\documentclass{PoS}

\title{Recurrent novae and long-term evolution of mass-accreting
white dwarfs -- toward the accurate mass retention efficiency}

\ShortTitle{RN and long-term evolution}

\author{\speaker{Mariko Kato}\thanks{Dept. of Astronomy, 
Keio University, Hiyoshi, Kouhoku-ku, Yokohama 223-8521, Japan}\\
        Keio University\\
        E-mail: \email{mariko.kato@hc.st.keio.ac.jp}}

\author{Izumi Hachisu\\
        University of Tokyo\\}

\author{Hideyuki Saio\\
        Tohoku University\\}

\abstract{
The mass growth rate of mass-accreting white dwarfs (WDs) is a key factor 
in binary evolution scenarios toward Type Ia supernovae.
Many authors have reported very different WD mass increasing rates. 
In this review, we clarify the reasons for such divergence,
some of which come from a lack of numerical techniques,
usage of old opacities, different assumptions for binary configurations,
inadequate initial conditions, and unrealistic mass-loss mechanisms.
We emphasize that these assumptions should be carefully 
chosen in calculating the 
long-term evolution of accreting WDs. Importantly, the mass-loss mechanism is 
the key process determining the mass retention efficiency: 
the best approach involves correctly incorporating the optically thick wind
because it is supported by the multiwavelength light curves of novae.
}

\FullConference{The Golden Age of Cataclysmic Variables 
and Related Objects IV\\
		11-16 September, 2017\\
		Palermo, Italy}

\begin{document}

\section{Introduction} \label{sec_introduction}
The mass retention efficiency of mass-accreting white dwarfs (WDs)
plays an important role in binary evolution scenarios toward 
Type Ia supernovae (SNe Ia) both for the single degenerate scenario
(SD: the progenitor is a binary consisting of a WD and non-degenerate star)
and double degenerate scenario (DD: a binary of two WDs).
In the SD scenario, the mass retention efficiency of WDs essentially governs 
the long-term evolution of the binary. 
In the DD scenario, the accreting WD grows in mass 
before reaching the second Roche-lobe overflow, thus, its growth rate 
affects the mass distribution of DD systems \cite{diS10}. 

Long-term evolutions of binaries including
a mass-accreting WD have been presented by many authors, but 
different authors have obtained very different WD mass increasing rates,
producing a diversity of SN Ia rates (e.g., \cite{bou13}).
These differences come from a variety of different assumptions 
for mass-accreting WDs, e.g.,
different adopted mass-loss formulae, different binary configurations,
numerical inaccuracies, and different opacities.
As the mass retention efficiency strongly affects binary evolution scenarios, 
here, we compare various numerical calculations, identify their drawbacks, 
and clarify the reasons for such divergence.

\begin{figure}
\includegraphics[width=.9\textwidth]{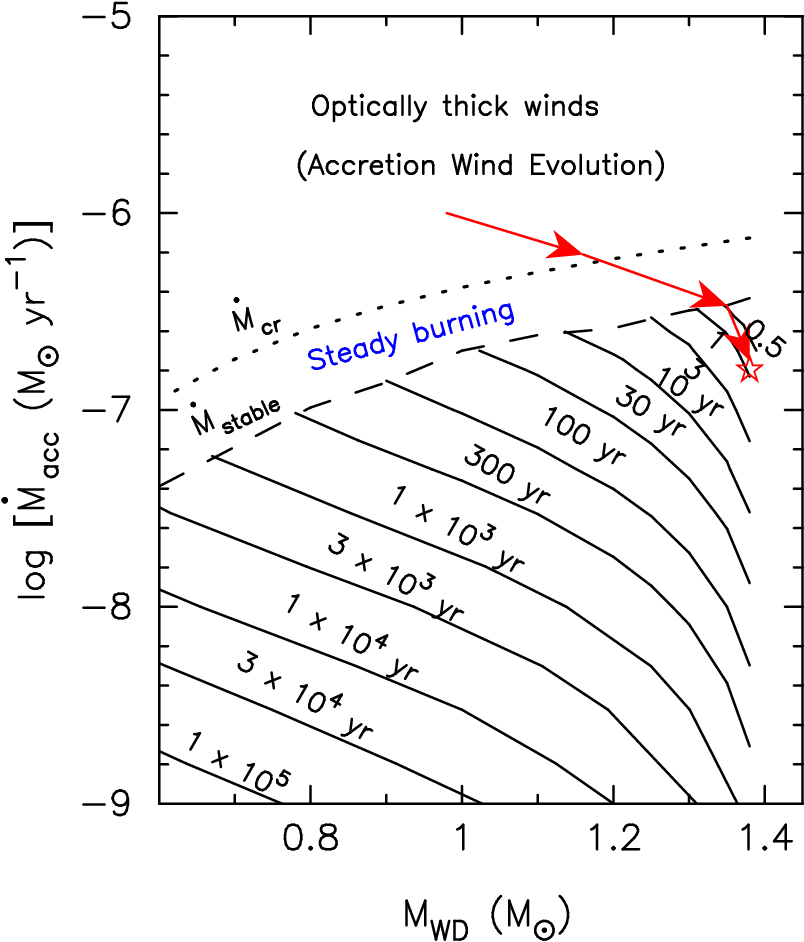}
\caption{
Response of WD envelopes to mass accretion on a WD mass versus
mass accretion rate diagram (so-called Nomoto diagram). 
The chemical composition of accreted matter is $X=0.7$, $Y=0.28$, and $Z=0.02$.
Above the stability line (dashed line labeled $\dot M_{\rm stable}$), 
hydrogen shell burning is stable.
Below this line, periodic shell flashes occur. 
The loci of the equi-recurrence period of novae are also plotted together
with its recurrence period. 
Above the dotted line labeled $\dot M_{\rm cr}$,
optically thick winds blow (labeled ``Accretion wind evolution'').
The periodic supersoft X-ray source (SSS) 
RX J0513-69 corresponds to this region. 
In the region between the two lines  
($\dot M_{\rm stable} \le \dot M_{\rm acc} \le \dot M_{\rm cr}$), we have
steady hydrogen shell burning with no optically thick winds.
Persistent SSSs are located in this region. 
A typical evolutionary path of SN Ia progenitors is indicated by the 
red solid line with arrows. The star mark indicates the position of 
the model of a nova in M31 with a one-year recurrence period
(M31N 2008-12a) \cite{kat17sha}, i.e., 1.38 $M_\odot$ and 
$1.6 \times 10^{-7}~M_\odot$yr$^{-1}$. 
\label{nomotoD}}
\end{figure}

\section{Physics of mass-accreting WDs}

\subsection{Stability} \label{sec_nomotodiagram}

Figure \ref{nomotoD} shows the response of WDs against various
mass accretion rates on a WD mass vs. mass accretion rate diagram
(so-called Nomoto diagram). Here, we consider the accretion of 
hydrogen-rich matter (solar abundance). 
The dashed line labeled $\dot M_{\rm stable}$ is the stability line
of hydrogen shell burning \cite{kat14shn,nom07,wol13}.
Below this line, i.e., $\dot M_{\rm acc} < \dot M_{\rm stable}$,
hydrogen nuclear burning is unstable and no steady burning can occur.
The WD suffers intermittent shell flashes, i.e., periodic nova outbursts.
In the upper region of this stability line, nuclear burning is stable.

\begin{figure}[]
\begin{center}
\includegraphics[width=.65\textwidth]{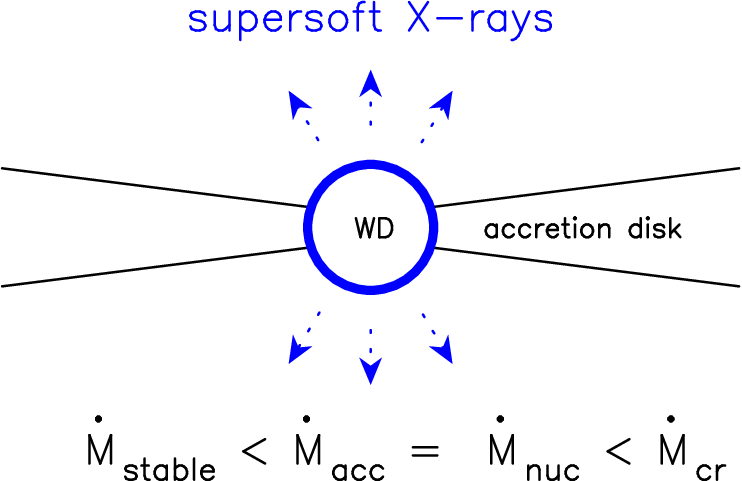}
\caption{
Schematic of steady hydrogen shell burning.
The WD accretes matter from an accretion disk
and burns hydrogen at the same rate as the accretion.
Such a configuration is often referred to as
a persistent SSS.
Taken from \cite{hac16forced}.}
\label{sss}
\end{center}
\end{figure}

\begin{figure}[]
\begin{center}
\includegraphics[width=.65\textwidth]{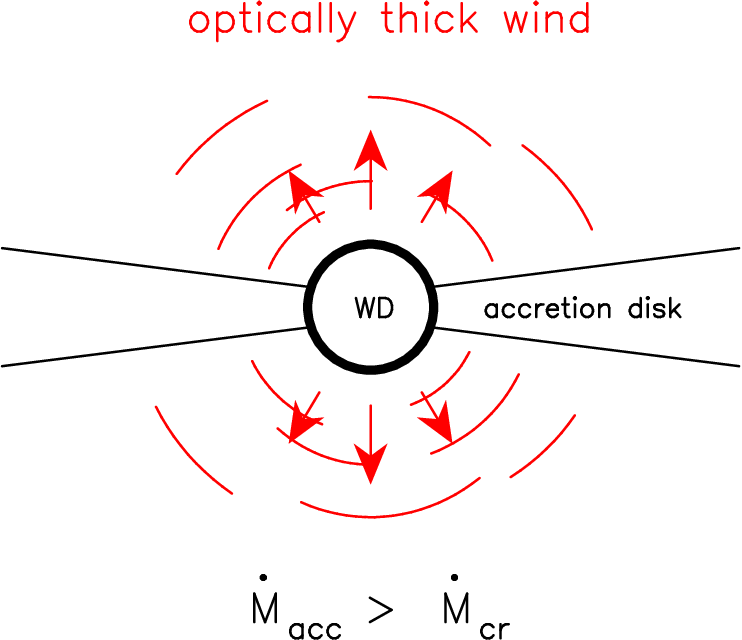}
\caption{
Same as in Figure 2, but for accretion wind evolution.
The WD accretes matter
from a disk and blows excess matter into the wind
($\dot M_{\rm wind}= \dot M_{\rm acc} - \dot M_{\rm cr}$).
Hydrogen steadily burns on the WD. Taken from \cite{hac16forced}.}
\label{windevolution}
\end{center}
\end{figure}

In the region between the two lines,
$\dot M_{\rm stable} \le \dot M_{\rm acc} \le \dot M_{\rm cr}$, we have
steady hydrogen shell burning with no optically thick winds.
All the accreted matter is burned into helium at the same rate as accretion. 
Thus, all the matter is accumulated on the WD (Figure \ref{sss}). 
The photospheric temperature of the WD envelope
is sufficiently high to emit supersoft X-rays.
Such objects are observed as persistent supersoft X-ray sources
(SSSs) \cite{van92,li97}.

In the region above the line of $\dot M_{\rm cr}$
($\dot M_{\rm acc} > \dot M_{\rm cr}$),
the WD accretes matter at a higher rate than the consumption rate of hydrogen burning.
The envelope expands and emits optically thick wind. 
Thus, the WD accretes matter via the accretion disk and, at the same time,
part of the accreted matter is blown in the other directions,
as illustrated in Figure \ref{windevolution}.
Hachisu \& Kato (\cite{hac03kb,hac03kc}) claimed 
that such accretion winds are present in the SSSs V Sge and RX J0513.9-6951. 
Their light curve models reasonably reproduce the cyclic behavior of the
optical/X-ray light curves with the cyclic on/off behavior of 
the accretion winds.  Such binaries are considered as objects 
corresponding to the accretion wind phase.
Thus, the three regions have corresponding objects.

Figure \ref{nomotoD} also shows a typical evolution path of the progenitor
in the modern SD scenario (Hachisu et al. 1999 \cite{hkn99,hknu99}).
The binary evolves from the optically thick wind phase,
through the SSS phase (steady H-burning phase)
and recurrent nova phase (H-shell flash phase), and
finally explodes as a SN Ia at the star mark.
This evolution path is realized only when optically thick wind is included. 
Small angular momentum that the winds carry away stabilizes 
the binary evolution and prevents the binary from evolving 
into a second common envelope stage (Hachisu et al. 1999a \cite{hkn99}). 

In contrast, many DD scenarios do not allow optically thick wind to blow,
because the original DD scenario is proposed before the OPAL opacity 
(see Section \ref{sec_opacity})
and the angular momentum loss due to the wind is not taken into account.
In such DD scenarios, many binaries are supposed to become a DD system after
the second common envelope evolution.
In other words, the main difference between the SD and DD scenarios are in the
accretion wind evolution (for more detail, see the review by Kato \& Hachisu \cite{kat12Review}).

\begin{figure}
\includegraphics[width=.93\textwidth]{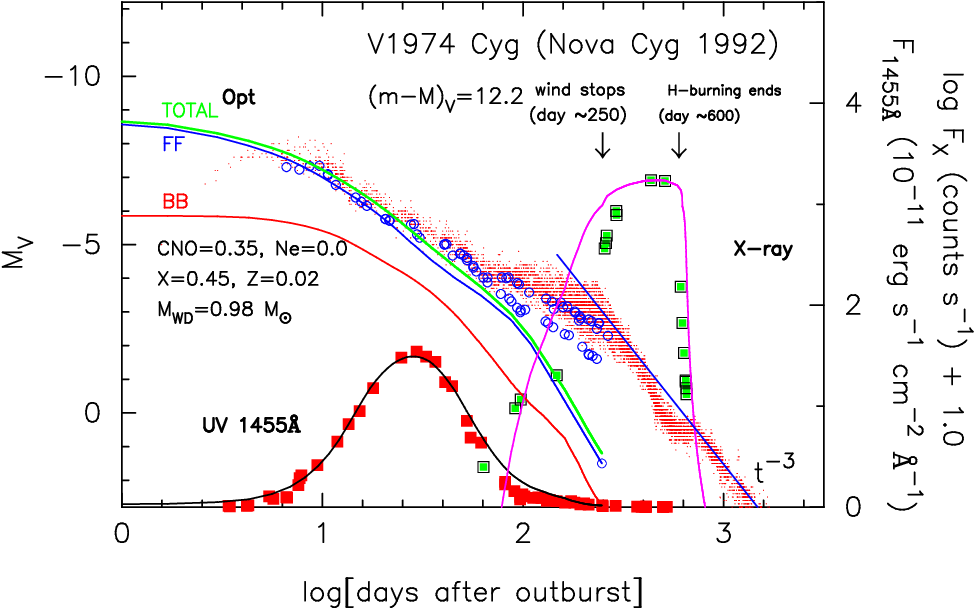}
\caption{
$V$ (blue open circles), visual (red dots), 
UV 1455 \AA ~(filled red squares), and supersoft X-ray (filled green squares
encircled by black squares) fluxes of the classical nova V1974 Cyg. 
The solid lines represent the theoretical multiwavelength light curves
of $M_{\rm WD}=0.98~M_\odot$
with the chemical composition of $X=0.45$, $Y=0.18$, $X_{\rm CNO}=0.35$,
and $Z=0.02$. The distance modulus in the $V$ band is $(m-M)_{\rm V}=12.2$.
The total optical flux (green line) is the summation of the contribution of
the free-free emission (blue line) from the outflowing winds 
and the blackbody emission (red line) at the photosphere. 
The $t^{-3}$ line indicates the trend of free-free flux
for a freely expanding nebula with no mass supply.
The optically thick wind stops at day $\sim 250$ and hydrogen burning
extinguishes at day $\sim 600$. This figure is reproduced from data 
in Hachisu \& Kato (2016) \cite{hac16k}.
}
\label{v1974cyg}
\end{figure}

\subsection{Physics of mass loss}\label{sec_ml}

During the evolution toward SN Ia explosion, binary WDs suffer mass loss
both in the accretion wind evolution and recurrent nova phase.
As the mass increasing rates of WDs depend strongly on the mass-loss rates,
we should adopt a mass-loss mechanism that is supported by observations.
The evolution of a nova outburst can be computed using the optically 
thick wind theory (Kato \& Hachisu 1994 \cite{kat94h}), and the 
light curves can be calculated from the obtained wind mass-loss rates 
(see e.g., Hachisu \& Kato 2006 \cite{hac06kb}; 
Kato 2012 \cite{kat12.palermo12}). The theoretically 
calculated light curves can reflect the observation, which in turn  
supports that optically thick winds are present in nova outbursts.

Figure \ref{v1974cyg} shows an example of light curve fits taken from  
Hachisu \& Kato (2016) \cite{hac16k}.
The duration of the optically bright phase is governed
by the mass-loss rate, which depends strongly on the WD mass and
weakly on the chemical composition of the envelope.
The supersoft X-ray phase also depends, although differently, on this WD mass and 
chemical composition. 
Thus, we calculated many light curves with different values of these two, and
selected a model that fit best the observed multiwavelength light curves.
The optical spectra of novae are basically free-free emission, so the 
optical light curve
is calculated using the wind mass-loss rate of the optically thick wind.
The ultraviolet (narrow 1455 \AA~ band) (black line)
and supersoft X-ray (magenta line) fluxes are
calculated assuming blackbody emission from the photosphere.
Thus, the optically thick wind successfully reproduces a number
of nova light curves (e.g., \cite{hac06kb,hac07,hac10k,hac15k,hac16k}).
The WD mass, distance, and some other properties can be obtained from this fit.
These are consistent with those estimated from other independent methods.
Thus, we can say that the mass-loss rate of the optically thick
wind is supported by many nova observations for a wide range of WD masses.

\subsection{Super-Eddington luminosity during nova outbursts}
\label{sec_superE}
Many novae have a super-Eddington phase in their early stages.
It can last for several days or more and its peak luminosity
often exceeds the Eddington limit by a factor of a few to several.
The physics of the ``super-Eddington'' stage is often misunderstood,
as described later in Section \ref{MESA}.
Thus, we explain here the mechanism of the super-Eddington luminosity.

The classical nova V1974 Cyg has a super-Eddington phase, of which
the peak luminosity exceeds the Eddington luminosity by 1.7 mag
and the duration lasts about 17 days.
Its optical light curve is well reproduced by a free-free emission light curve
 (see Figure \ref{v1974cyg}).
Note that the optical emission from the blackbody photosphere 
(red line labeled ``BB'') is slightly sub-Eddington,
while the optical free-free emission reaches super-Eddington luminosities
(blue line labeled ``FF''). Free-free emission comes from the optically thin plasma 
outside the photosphere (for details see \cite{hac16k}).

The emissivity of the free-free emission is
proportional to the square of the wind mass-loss rate, i.e.,
$F_\nu \propto \dot M_{\rm wind}^2$.
The free-free flux can exceed the Eddington value 
when the wind mass-loss rate is sufficiently high in the early stage.
The mass-loss rate decreases as the photospheric temperature
increases with time, so the free-free emission light curve decays with time.
In this case, the optically thick winds successfully reproduce a number of
nova light curves including early super-Eddington phases.

\begin{figure}
\includegraphics[width=.45\textwidth]{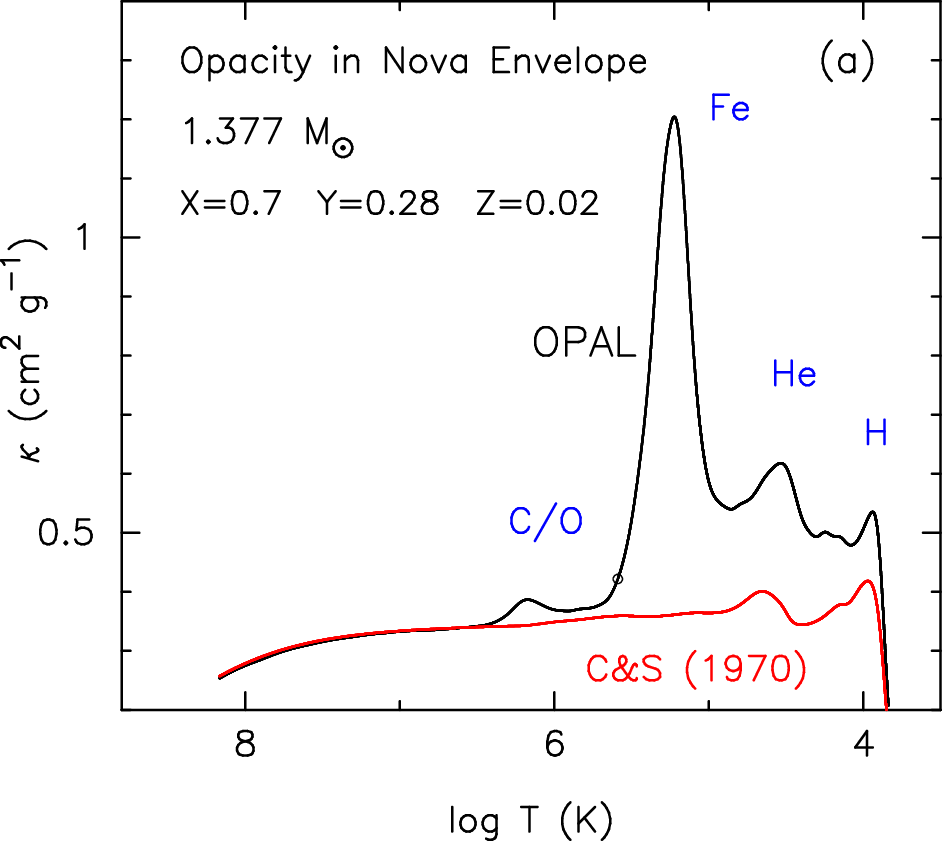}
\includegraphics[width=.55\textwidth]{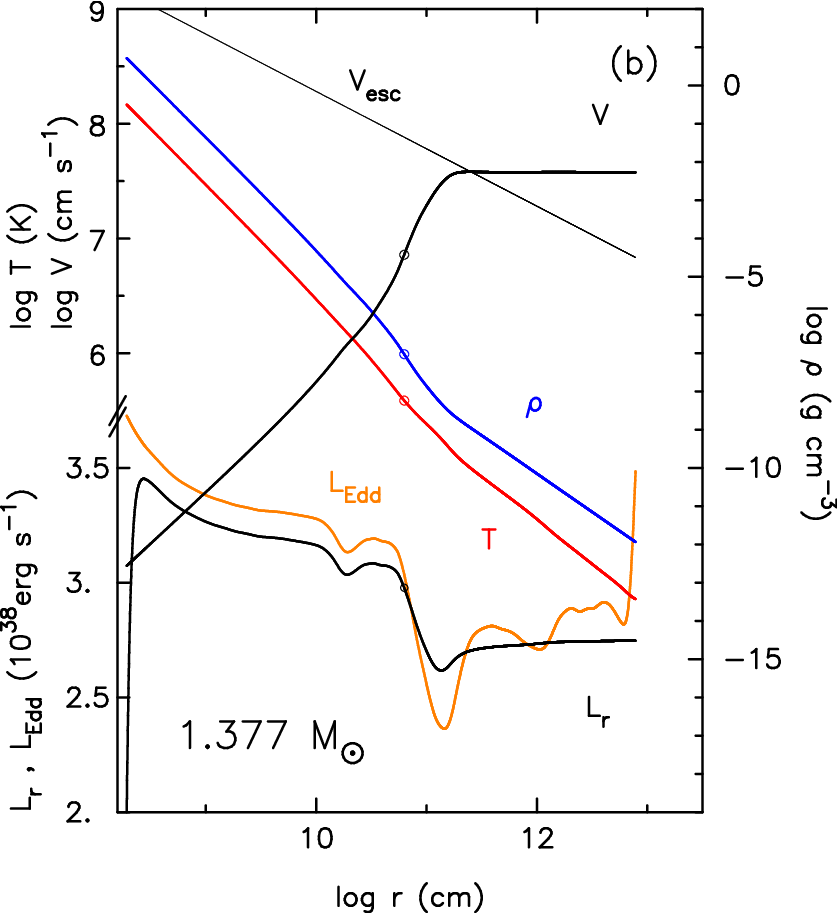}
\caption{
(a) OPAL opacity distribution (black line) of a wind mass-loss solution 
for a 1.377 $M_\odot$ WD with a chemical composition of
$X=0.7$, $Y=0.28$, and $X=0.02$.  The red line represents the Los Alamos
opacity (Cox and Stewart (1970) \cite{cox70a,cox70b}) for the temperature
and density of the solution denoted by the black line.
The rightmost end corresponds to the photosphere.  
The small open circle denotes the sonic point (the critical point 
in \cite{kat94h}).
(b) Envelope structure of the same solution. 
The rightmost edge of each line corresponds to the photosphere.
}
\label{opac}
\end{figure}

In the HR diagram the bolometric luminosity $L_{\rm ph}$ is almost
constant in the bright phase and does not exceed the Eddington luminosity of 
electron scattering (see Figure \ref{hr} below).
This property can be seen in any other papers that show HR diagrams for one
nova outburst cycle
[e.g., \cite{ibe82,kov98,den13,kat17sha,kat17shb}], being 
independent of the old/new opacities.
This is because the large nuclear luminosity produced 
in the nuclear burning region is
quickly absorbed in a layer around the nuclear burning zone and, as a result,
the outward photon flux is sub-Eddington 
(see e.g., \cite{kat17sha,kat17shb} for detail).

In contrast, Prialnik \& Kovetz (1995) \cite{pri95} and 
Yaron et al. (2005) \cite{yar05} reported super-Eddington luminosities 
by a factor of several to up to a few tens
(against the Eddington luminosity for electron scattering opacity).
This excess is larger than the possible error of a factor of 2 
in luminosity (Prialnik \& Kovetz 1995 \cite{pri95}).

There are no hint for the mechanism of such super-Eddington luminosity
in their papers.   We encourage the authors to publish the envelope 
structures, photospheric temperature and luminosity and explain 
why the super-Eddington luminosity is realized. 
In Section \ref{sec_masszone}, we discuss that sufficient number 
of mass grids is necessary to obtain accurate numerical solutions.
Considering differences from many other calculations cited above, 
we suggest the authors to confirm their numerical results 
by solving the envelope structure up to the photosphere 
with sufficient mass grids.

Shaviv (2001) \cite{sha01} presented the idea of reduced opacity
in a porous envelope for the super-Eddington luminosity mechanism.
This porous instability, however, has not yet been studied
for realistic cases of the nova envelope. Hence, we
do not know whether the super-Eddington luminosity
can be realized with this instability
(for detailed discussion, see \cite{kat12.palermo12}).
As discussed above, the observed light curves of novae
can be well reproduced without this mechanism.

\section{Pitfalls in numerical calculation} \label{sec_tech}

In this section, we describe two points that clearly give very
different numerical results.  One is the opacity and the other is
the number of mass shell grids.

\subsection{Opacity}\label{sec_opacity}

When the radiative opacities were recalculated at the beginning of the 1990s,
there was a drastic development in theoretical nova research.
The revised opacity tables showed a very large enhancement 
at $\log T$ (K)$ \sim 5.2$ owing to the
Fe line transitions (OPAL opacity: \cite{igl87,rog92,igl96}, 
and OP opacity: \cite{sea94}),
which was not included in the Los Alamos opacities \cite{cox70a,cox70b}.
The large opacity peak changes the nova envelope structure, especially
when the luminosity approaches the Eddington luminosity. 
Strong winds are accelerated and, as a result, 
the timescales of nova outbursts are drastically reduced. 
The distribution of the nova speed class, for the first time, agreed well with 
the mass distribution of the WD, 
i.e., massive WDs ($> 1~M_\odot$) correspond to fast novae 
and less massive WDs ($0.6-0.7~M_\odot$)
fit slow novae \cite{kat94h,pri95}.
Moreover, we reached an overall understanding of nova evolution, 
not only qualitatively but also quantitatively.
The optical/IR light curves were calculated based on the optically thick wind
solutions and their evolution of the wind mass-loss rates now 
explain effectively the observed light curves. Multi-wavelength 
light curve fittings of novae now quantitatively agree with 
the light curves of a number of classical novae, as explained in
the previous section.

Figure \ref{opac}a presents the opacity distribution
through an envelope of an optically thick
wind solution corresponding to a very extended nova outburst stage
on a $1.377~M_\odot$ WD calculated with the OPAL opacity
for the solar composition, $X=0.7$, $Y=0.28$, and $Z=0.02$ \cite{kat94h}.
Figure \ref{opac}b shows the internal structure: velocity distribution, 
density, temperature, diffusive luminosity, 
and the local Eddington luminosity, which is defined as
\begin{equation}
L_{\rm Edd} = {4\pi cG{M_{\rm WD}} \over\kappa},
\label{equation_Edd}
\end{equation}
where $\kappa$ is the opacity; we used the OPAL opacity.
As $\kappa$ is a local variable as a function of
the temperature and density, the local Eddington luminosity
varies significantly with the radius $r$.
Two super-Eddington regions appear 
at $\log r$ (cm) $\sim 11.2$ and 12.1, corresponding
to the opacity peaks due to the Fe and He ionization regions, respectively.
The sonic point (open circles, critical point, see \cite{kat94h}) 
appears at $\log r$ (cm)= 10.80.
The wind is accelerated around this point and reaches 
a terminal velocity far below the photosphere.

Figure \ref{opac}a also shows the Los Alamos opacity
(red line labeled ``C\&S (1970)''), calculated for the
temperature and density of the wind solution in Figure \ref{opac}b.
Here, we used Iben's analytical opacity formula \cite{iben75} 
that represents the Los Alamos opacity \cite{cox70a,cox70b}.
This old opacity does not exhibit the Fe opacity peak, so it does not
accelerate strong winds and, thus, 
the nova evolution timescales are extremely long
(see, e.g., Figure 18 of \cite{kat94h}).

Currently, the new opacities are widely used in stellar
evolution calculations. Especially, in a nova outburst, 
these opacities are essential because the envelope structure
is substantially affected 
when the luminosity is close to the Eddington luminosity.
Several researchers switched to using the new opacities in the 1990s.
For each group, the first papers using the OPAL opacity were
Kato \& Hachisu (1994) \cite{kat94h}, Prialnik \& Kovetz (1995) 
\cite{pri95}, and Starrfield et al. (1998) \cite{sta98}.

Some people, including Cassisi, Iben, and Tutukov, did not adopt
the new opacities and continued using the old opacity.
It should be noted that Cassisi et al. (1998) \cite{cassi98} emphasized that
'the Los Alamos opacities are very similar to the OPAL opacities'
(see the last sentence of Section 4 in their paper).
This statement is clearly incorrect, as shown in Figure \ref{opac}a.
Piersanti, Tornamb\'e, \& Yungelson had long been users of the
Los Alamos opacity, but in a recent paper, they adopted the OPAL
opacity (Piersanti, Tornamb\'e, \& Yungelson 2014 \cite{pie14}).
It is not easy to check which opacity is adopted, because not all the papers
specify their opacity, especially those with the old opacity.

We encourage authors to mention the adopted opacity in each paper.
In addition, readers should carefully check what opacities are adopted
and be aware that if the old opacities are used, their numerical calculation could
lead to very different conclusions for binary evolution.

\begin{table}
\begin{tabular}{lllllll}
\hline
Authors  & Mass Grids  &Flash Cycles & H/He  &Code & Mass Loss\\ \hline
Starrfield et al. (1998) & 95 & $< 1$ &H & NOVA & -- \\
Starrfield et al. (2012) & 400   & $< 1$ & H &  NOVA & --  \\
Newsham et al. (2014) &unknown&  18  & H & MESA & similar to Eq (\ref{equation_dmdt})\\
Prialnik \& Kovetz (1995) & $200-300^{a}$ & 8 &  H & Prialnik & Eq (\ref{eq:dmdt}) \\ 
Yaron et al. (2005) & $200-300^{a}$ & $<$ a few & H&Prialnik& Eq (\ref{eq:dmdt})\\
Epelstain et al. (2007) & 200-400   & 1000    &H &Prialnik& Eq (\ref{eq:dmdt}) \\
Hillman et al. (2015) & a few 100 & 36000 & H & Prialnik & Eq (\ref{eq:dmdt})  \\
Hillman et al. (2016) & a few 100$^{a,b}$ &  400& He &Prialnik & Eq (\ref{eq:dmdt})   \\
Kovetz (1998) & $< 60^{c}$  & 2    & H & Prialnik & steady wind\\
Idan et al. (2013) &  8000 & 4000& H &Idan & steady wind\\
Kato et al. (2017b)& $4000^{d}$ & 1500 & H & Saio & steady wind$^{e}$\\
Denissenkov et al. (2013) &  1000-2000 &  4 &H& MESA & Eq (\ref{equation_dmdt})\\
Ma et al. (2013)  & 2000  & 10 & H& MESA & modified Eq (\ref{equation_dmdt})\\
Wang et al. (2015a)  & unknown & 18 & He & MESA &modified Eq (\ref{equation_dmdt})\\
Wu et al. (2017)  & 1600$^{f}$  & 1000 & He & MESA &modified Eq (\ref{equation_dmdt})\\
\hline
\end{tabular}
\caption{
Comparison of various numerical calculations. 
(a) Private communication with Prialnik (2017). 
(b) Less than 5 grids in the entire He-rich region after 40 flash cycles 
(see Figure 6).
(c) For the wind region.
(d) $> 2000$ mass grids for interior and about 2000 for wind region.
(e) The mass loss rate is determined from fitting with a wind solution
(see Section 4.4).
(f) Private communication with Wu (2017).}
\label{table_mesh}
\end{table}

\subsection{The number of mass shell grids}
\label{sec_masszone}
In the pioneering era of nova calculation, 
Nariai, Nomoto, \& Sugimoto (1980) \cite{nar80} warned 
that a small number of mass shell grids could 
produce what seemed to be correct. 
In their test calculation with coarse mass zoning,
the calculation converged to a model that had 
a very different structure with 
less extension in $\Delta \ln r$ and $\Delta \ln P$ of the envelope. 
Thus, the expansion velocity would be low.
This apparently non-diagnostic model is, however, an
artifact because of the small number of mass grids.
Such models tend to result in low expansion velocities, that is, less
violent explosions, resulting in smaller masses of ejecta. As a result,
it produces larger WD mass increasing rates. 

Table \ref{table_mesh} lists representative shell flash calculations
that adopt the OPAL opacity. 
Calculations with the old opacities are not included in this table.
From top to bottom, there are three entries for Starrfield's group,
7 for Israeli code users, one for the Japanese group, 
followed by four MESA code users.
The list also shows the typical number of mass grids,
number of flash cycles, nuclear fuel of the flash, numerical code,
and adopted mass-loss formula.
For the papers including several different cases, we adopt the model
described in most detail or the model of more massive WDs.
The mass-loss algorithm will be discussed in Section \ref{sec_MLalgorithm}.

\begin{figure}
\begin{center}
\includegraphics[width=.85\textwidth]{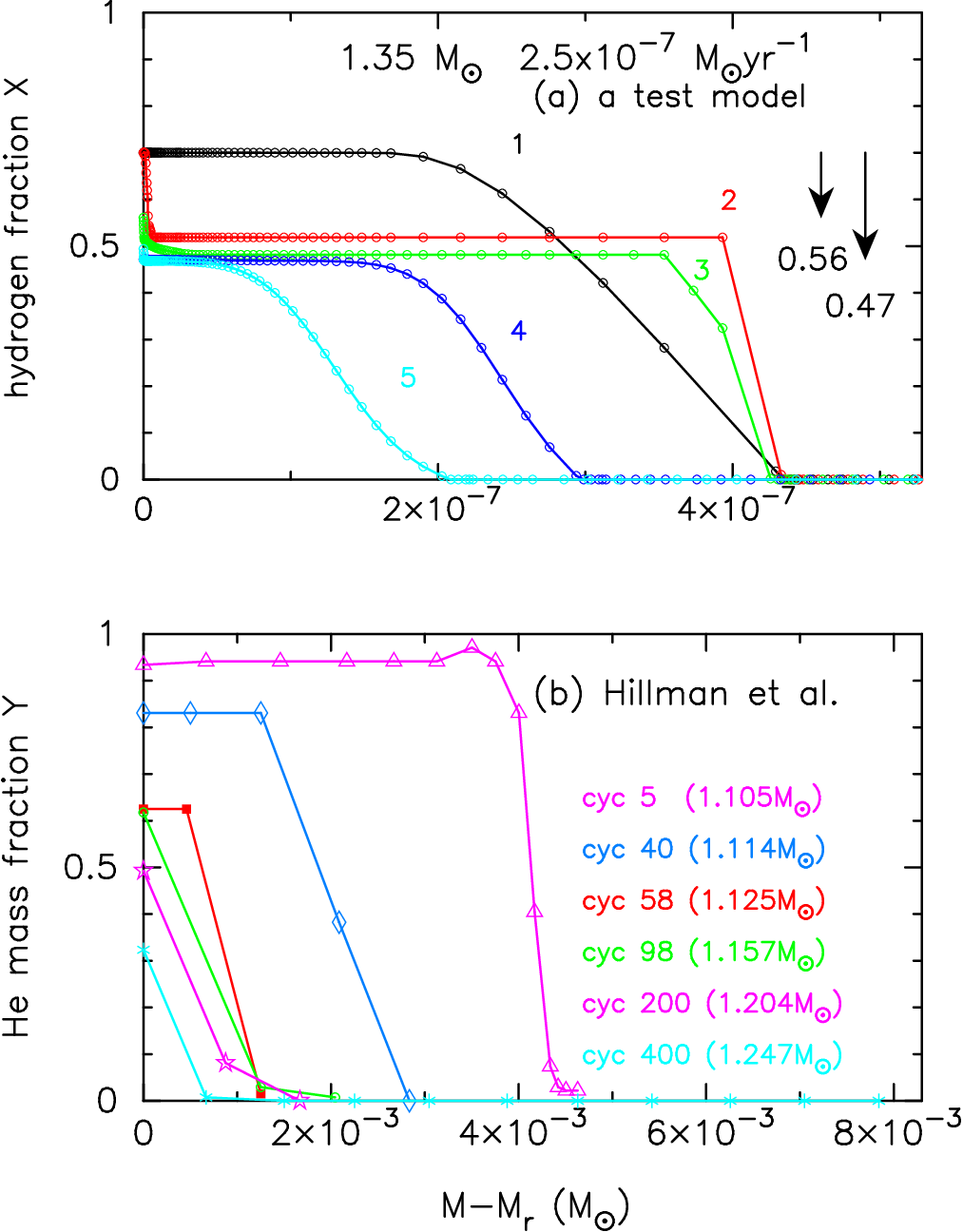}
\caption{  
(a) Temporal change in hydrogen mass fraction in the envelope 
of a $1.35~M_\odot$ WD with hydrogen-rich mass accretion
rate of $ 2.5 \times 10^{-7}~M_\odot$yr$^{-1}$ ($P_{\rm rec}=1.1$ yr)
for a test model with coarse mass zoning.
Stage 1: onset of thermonuclear runaway.
Stage 2: shortly after ignition. Convection widely develops in 
the flat portion 
and $X$ jumps from zero to 0.52 in one mesh. 
Stage 3: because of the re-zoning process, 
the one-mesh jump is partly restored. 
Stage 4: wind mass loss occurs. The surface region is blown in the wind,
so the boundary of the H and He layers shifts leftward. 
Stage 5: the shell flash ends and accretion restarts.
The downward arrows indicate the decrease in hydrogen fraction.
It decreases down to $X=0.47$ for the coarse zoning model, while 
the hydrogen decreases to $X=0.56$ for the model with 
a sufficient number of mass grids (not shown here).
(b) Helium mass fraction for six different cycles of
Hillman et al. (2016)\cite{hil16}.   
The WD mass is indicated after the flash cycle numbers
in parenthesis. The He mass accretion rate is the one corresponding to
the hydrogen mass accretion rate of $2 \times 10^{-7}~M_\odot$yr$^{-1}$.
There is no mass ejection after cycle 98. 
The helium mass fraction significantly decreases through 395 cycles.
Reproduction from the data in Figure 9 
of Hillman et al. (2016)\cite{hil16}. 
}
\label{QXm135}
\end{center}
\end{figure}

To demonstrate the importance of a sufficient number of mass shell grids, 
we calculated a test hydrogen shell flash model on a $1.35~M_\odot$ WD
with coarse mass grids 
for a mass accretion rate of $2.5 \times 10^{-7}~M_\odot$yr$^{-1}$
(in Figure \ref{QXm135}a).
The total grid number is about 550.
The hydrogen profile of the selected stages is represented by thick lines with
dots that indicate each mass grid. The hydrogen shell flash starts
at stage 1.
Convection occurs to widely mix the pre-existing helium-rich layer
($2 \times 10^{-7}M_\odot < M-M_r < 4.4 \times 10^{-7}M_\odot$)
as well as the freshly produced helium.
The uppermost layer is radiative, where the convection cannot reach,
so the hydrogen content maintains the
original value of $X=0.7$.
As the mass zoning is sparse, the hydrogen content jumps from zero to 0.5
within the neighboring one/two meshes in stages 2 and 3.
We adopt a re-zoning technique,
so this coarse mass zoning area disappears in the later phases,
as shown in stages 3 and 4.
The hydrogen content deceases to $X=0.47$.
In our other test model with sufficient mass grids (total $\sim 1000$ grids),
the hydrogen decrease stopped at $X=0.56$.
This difference of $\Delta X=0.09$ is not physical but numerical
due to the insufficient resolution ($\sim 550$ grids).  
Thus, a small number of grids leads to an artificial decrease
of hydrogen content even for only one cycle of outburst.  
This artificial difference could be accumulated for a number 
of outburst cycles. 

Figure \ref{QXm135}b shows the temporal change in He mass fraction 
in the envelope accreting He in successive
helium flashes calculated by Hillman et al. (2016) \cite{hil16}. 
Each line corresponds to a different cycle. 
The phase of each cycle is not specified in their paper. 
The helium mass fraction decreases with the cycle number.
In and after cycle 40, there is a large jump 
in helium mass fraction between one/two 
mass grids, resembling our test model in stages 2 and 3 in the upper figure.
The temperature at the outermost mass zone is extremely high ($> 10^8$ K;
beyond the upper boundary of their Figure 8).
Based on this calculation, Hillman et al. (2016) \cite{hil16} concluded
that helium shell flashes weaken in intensity and finally 
stabilize. Therefore, the mass retention efficiency of helium accretion becomes 100 \%. 

This conclusion, however, is obtained from the calculation 
with a very small number of mass grids in helium rich layers 
(only $\lesssim5$ mass grids after 40 flash cycles).
As the temperature is very high ($> 10^8$ K) in the outermost mass zone,
the freshly accreted helium matter is
instantly consumed by nuclear reactions, so the
helium mass fraction is much smaller than the original value ($Y = 0.98$) 
even in the outermost mass zone. 
This is probably the reason that their helium flashes become very weak even 
below the stability line of helium shell burning.

\begin{figure}
\includegraphics[width=.98\textwidth]{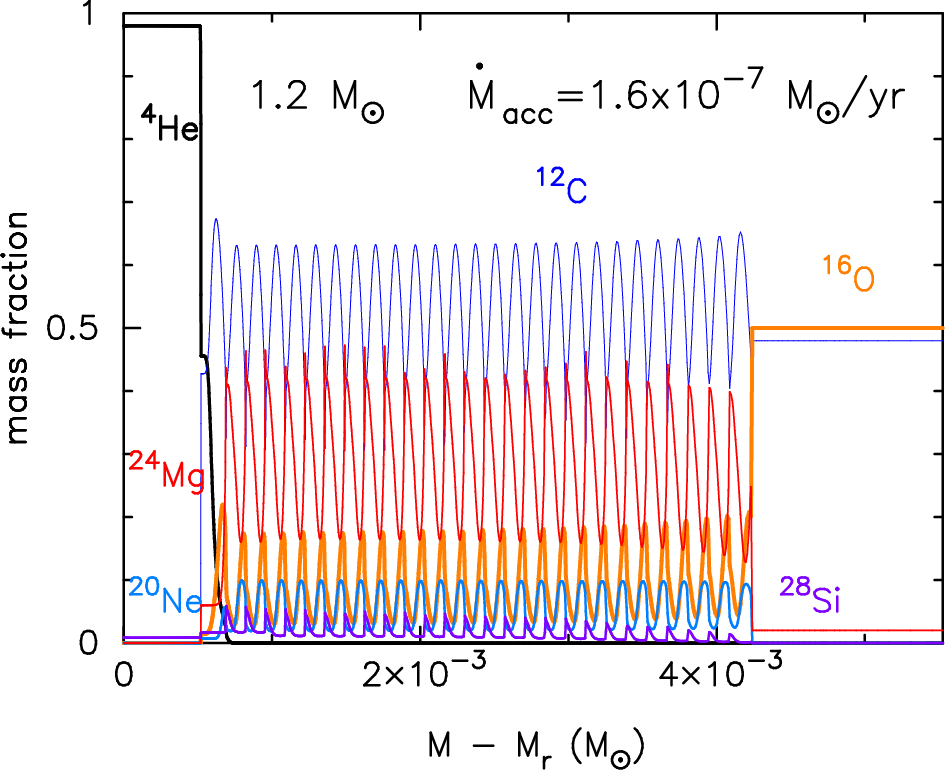}
\caption{Distributions of $^4$He, $^{12}$C, $^{16}$O, $^{20}$Ne, 
$^{24}$Mg and $^{28}$Si in the envelope of a $1.2~M_\odot$ WD  
with a helium mass accretion rate of
$\dot M_{\rm acc}=1.6 \times 10^{-7}~M_\odot$yr$^{-1}$ 
after it has undergone 27 helium shell flashes \cite{kat18sh}. 
} 
\label{m12qx.palermo}
\end{figure}

A good contrast is shown in the internal structure of our preliminary model
(Figure \ref{m12qx.palermo}) on a $1.2~M_\odot$ WD with a He mass 
accretion rate of $1.6 \times 10^{-7}~M_\odot$yr$^{-1}$ \cite{kat18sh}.
The mass fraction of various elements exhibits 27 peaks
corresponding to the cyclic temperature
changes during each He-shell flash because of
a quiescent phase after the WD experienced
27 successive helium shell flashes.
The total number of mass grids is 1628, 
of which 864 grids are allocated to the region
with 26 oscillations 
($6.6 \times 10^{-4}~M_\odot < M-M_r < 4.2 \times 10^{-3}~M_\odot$), 
396 grids are allocated to the outer region ($Y > 0.05$) including outermost 
cyclic peak of carbon.  
The next outburst occurs at the inner edge of He rich layer, i.e., 
around $Y=0.05$ ($ M-M_r \sim 6.6 \times 10^{-4}~M_\odot$), thus the 
helium profile is important for the calculation of the next outburst.  
Comparing Hillman et al.'s (2016) Figure 9 
(or present Figure \ref{QXm135}b) with Figure \ref{m12qx.palermo}, 
we conclude that (1) the mass included in the outermost mass zone
(after cycle 98) is comparable or larger than the ignition mass of 
helium shell flash for their helium mass-accretion rate. 
Unstable shell flashes do not occur because helium 
shell-burning is artificially stabilized when the envelope
mass already (or always) exceeds the ignition mass due to the coarse
mass zoning 
(see, e.g., Nomoto et al. 2017, for more detail description)\cite{nom07}.   
(2) Clearly a few hundred mass grids are not enough to accurately
resolve the elemental composition profiles after 400 shell flashes.

\begin{figure}
\includegraphics[width=.98\textwidth]{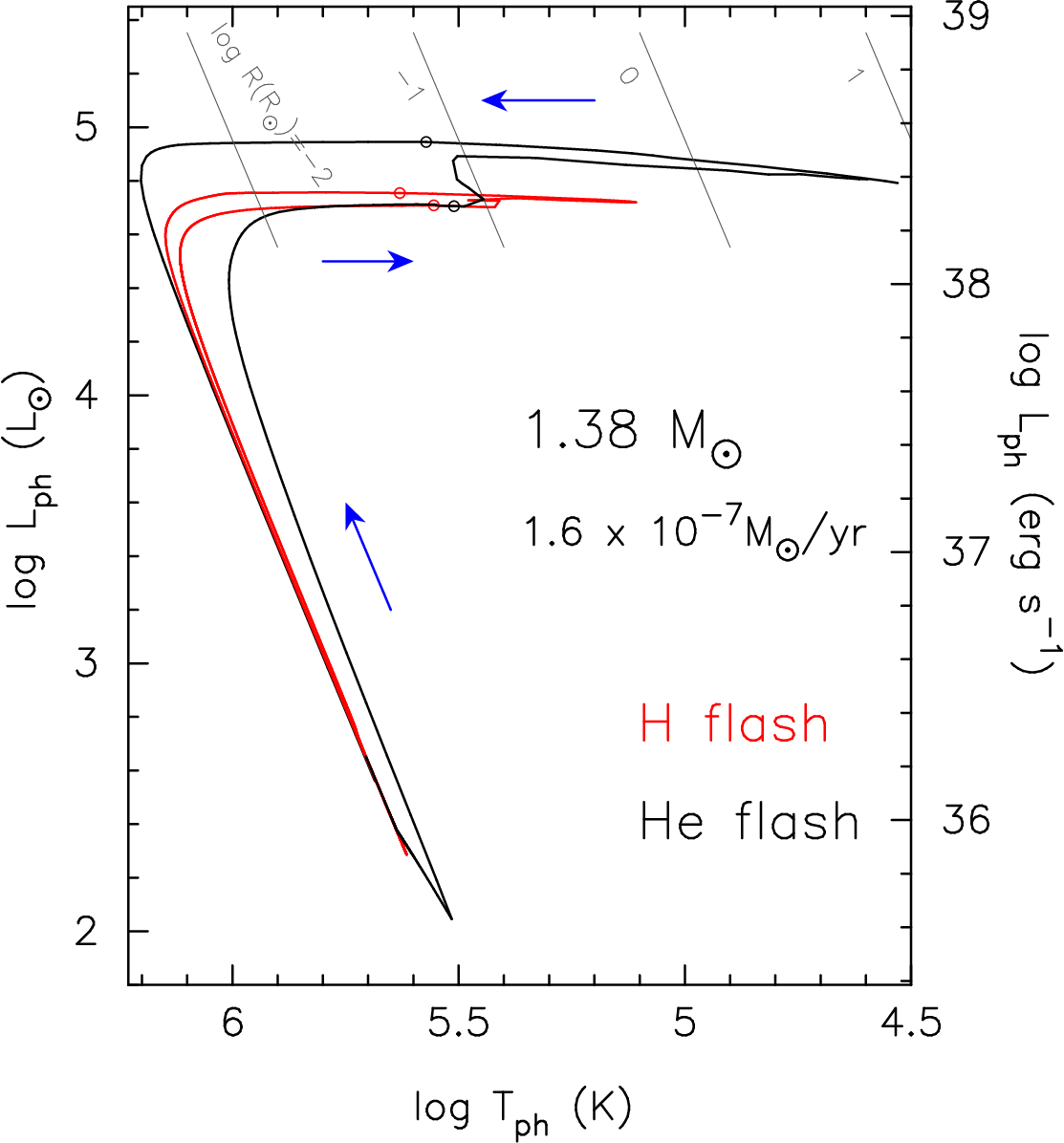}
\caption{Evolution track of a hydrogen shell flash (red line)
followed by a helium shell flash (black line).
The blue arrows indicate the direction of evolution.
The thin solid black lines indicate the lines of 
$\log R_{\rm ph}/R_\odot=-2$, $-1$, 0, and $+1$, from left to right. 
The optically thick wind blows 
in the region right of the small open circles of each track.
The small zig-zag path at $\log T$ (K) $\sim 5.5$ 
is due to the change in chemical composition, 
i.e., the surface layer with the original composition of 
accreted matter is blown away in the wind both for the H and He shell flashes.
The red line is taken from \cite{kat17sha} 
and the black line is taken from \cite{kat18sh}. 
The decay phase (leftward track) in the black line is 
approximated by a steady-state solution. 
}
\label{hr}
\end{figure}

Figure \ref{hr} presents a HR diagram for hydrogen and helium shell flashes
on a 1.38~$M_\odot$ WD with a mass accretion rate of 
$\dot M_{\rm acc}=1.6 \times 10^{-7}~M_\odot$yr$^{-1}$ \cite{kat17shb}. 
The red line represents the last hydrogen shell flash 
of the 1500 successive flash calculation, and the black line represents
the helium shell flash immediately after that.  
In both the tracks, 
optically thick wind mass-loss takes place when the photospheric
temperature decreases to $\log T$ (K) $\sim 5.5$ (small open circles).
There is a small zig-zag path at $\log T$ (K) $\sim 5.5$ in each track 
owing to the change in chemical composition in the surface layer. When the 
surface layer is blown in the wind, the hydrogen (helium) mass
fraction quickly decreases (see Figure \ref{QXm135}a), 
which causes a decrease in the opacity. Thus, the luminosity increases 
after a small zig-zag track caused by the structural change. 
If we adopt coarse mass zoning in the surface area, we will not follow 
such a zig-zag path. 

Starrfield et al. (1998) \cite{sta98} adopted a small grid number 
as in Table \ref{table_mesh}.
This small number caused confusion for the stability analysis,
which will be discussed in Section \ref{section_stability}.

\section{Mass-loss algorithm: assumed mechanism}\label{sec_MLalgorithm}

As is well known, Henyey-type codes, widely used 
in stellar evolution calculations,
are numerically difficult to calculate 
in the expanding phase of a nova outburst when the
envelope becomes radiation-dominant.
After the new opacities (OP and OPAL) appeared at the beginning of 1990,
the numerical calculations became much more difficult
because the density quickly
decreases at the low-temperature side of the opacity peak.
To continue the calculation beyond this stage,
various authors have adopted various numerical procedures that may represent
different mass-loss mechanisms.
There is no need to say that different mass-loss formulae result in different
mass increasing rates of WDs, thus, 
different fates of the long-term evolution of binaries must be calculated.
Here, we examine various mass-loss formulae
and determine the best one that represents a realistic mass-loss mechanism.
We explain four different treatments for mass-loss. 
The first three methods are those that avoid calculating the envelope region
affected by the large opacity peak and the last one calculates 
all the envelope regions including the lower-temperature side
of the opacity peak.

\subsection{Treatment by Prialnik's group}\label{sec_subtract}

In the pioneering work, Prialnik (1986) \cite{pri86} 
first calculated one nova outburst cycle (with 120 mass grids)
and presented the temporal change in the internal structure 
of an expanding envelope.
In her calculation,
the mass-loss rate, a time-dependent local variable,
\begin{equation}
\dot m = 4 \pi r^2 \rho v,
\label{eq:dmdt}
\end{equation}
is quite flat outside the hydrogen burning region at a later phase 
of the outburst. This implies that 
the envelope reaches a steady-state at a later phase.

Prialnik \& Kovetz (1995) adopted the following mass-loss algorithm
to continue the multicycle evolution calculation \cite{pri95}.
The occurrence of mass loss is detected when the supersonic
wind regime appears in the surface zone.
The mass-loss rate had been calculated from Equation (\ref{eq:dmdt})
at the bottom of the zone. 
The mass loss is treated by the subtraction of $\dot m \delta t$
from the outermost mass shell, where $\delta t$ is the time step.
The photospheric luminosity, temperature and radius were not
presented in their paper. 
The same computational code and mass-loss algorithm
are adopted in the same group \cite{yar05,epe07,hil15,hil16}.

A good way to know whether an adopted algorithm is consistent with the 
mass-loss of novae is to compare the model light curves with observations. 
Until recently, we were not able to compare their calculations with
observations because no multi-wavelength light curves were available.
Hillman et al. (2014) \cite{hil14} presented multi-wavelength light curves 
(visual, UV, X-ray magnitudes: with no definition given) based on the solution 
of Prialnik \& Kovetz (1995)\cite{pri95} 
and Yaron et al. (2005) \cite{yar05}. There are no description how they 
determined the effective temperature and radius from the temperature and 
radius at the outermost mass zone.
These light curves resemble a rectangular shape, i.e., all the three magnitudes
suddenly increase at the same time, maintain the peak values for some time,
and drop at the same time. Their effective temperature also changes
in a rectangular manner, i.e., a sudden decrease to a minimum value 
followed by a sudden rise.
These light curves do not resemble real nova light curves.
For example, V1974 Cyg (Figure \ref{v1974cyg}), a typical classical nova,
exhibits an optical peak followed by a UV dominant phase that is replaced by
a supersoft X-ray emission dominant phase.
In addition, a gradual temperature increase 
is reported in many novae, e.g., UV spectra indicate that higher ionization
lines reach a maximum flux at later times \cite{cas02,cas04,kat09v838her}.
Because the light curves of novae are controlled by mass-loss rates, 
Prialnik's mass-loss scheme may not represent realistic mass ejection
in nova outbursts.

\subsection{Assumption adopted by MESA code}\label{MESA}

One of the ways to avoid the numerical difficulties associated with the opacity peak is 
to assume some mass loss before the wide expansion of the envelope.
Denissenkov et al. (2013) \cite{den13} calculated multicycle nova outbursts with the MESA code,
assuming mass loss when the surface luminosity
reaches the Eddington luminosity at the surface,
\begin{equation}
L_{\rm Edd,s} = {4\pi cG{M_{\rm WD}} \over\kappa_{\rm s}},
\label{equation_Edds}
\end{equation}
where $\kappa_{\rm s}$ is the opacity at the surface.
The mass-loss rate (negative value) can be given by 
\begin{equation}
\dot M = -2 {{(L-L_{\rm Edd,s})}\over v_{\rm esc}^2},
\label{equation_dmdt}
\end{equation}
where $L$ is the luminosity, $v_{\rm esc}=\sqrt{2GM_{\rm WD}/R}$,
and $R$ is the surface radius.
This implies that the excess energy $L-L_{\rm Edd,s}$ is converted to
gravitational energy between the surface of the envelope and infinity.
In their model of a $1.15~M_\odot$ WD, the radius expands up to $3 R_\odot$
($\log T$ (K) $\sim 4.63$).
Apparently, the expansion stops at the opacity peak corresponding
to He ionization (see their Figure \ref{opac}).
Detailed information, such as the temporal change in the photospheric
temperature or wind mass-loss rate, has not been given.
Newsham et al. (2014) also used the MESA code with a similar assumption, but an exact definition
has not been provided in their proceedings paper \cite{new14}.

This assumption, that all the excess energy $(L-L_{\rm Edd,s})$ is converted into
kinetic energy, has not been physically confirmed. The Eddington luminosity 
is the maximum luminosity of a hydrostatic envelope of self-gravitating stars.
This assumption breaks down once wind mass loss takes place.
For example, in the steady wind, the Eddington luminosity in Figure \ref{opac}b
does not represent the upper limit of the luminosity.
In some extreme cases, Kato (1984) showed that
when the velocity gradient term $v dv/dr$ is non-negligible,
the luminosity exceeds the Eddington luminosity $(L_{\rm ph} > L_{\rm Edd,s})$,
even when the kinetic energy $v^2/2$ is small,
where $L_{\rm ph}$ is the luminosity at the photosphere \cite{kat83}.

Note, in steady-state winds, ``super-Eddington'' regions can appear,
as shown in Figure \ref{opac}b, where the luminosity exceeds the
local Eddington luminosity corresponding to the peak of the opacity.
In this case, however, the photospheric luminosity does not exceed 
the Eddington luminosity of electron scattering. 
The MESA code approximation is closely related to these local
(apparent) super-Eddington regions.  This is not the so-called 
``super-Eddington luminosity'' in which the observed total flux
greatly exceeds the Eddington luminosity of electron-scattering
opacity, similar to in Figure \ref{v1974cyg}.  

Some authors  \cite{ma13,wan15a,wu17} called the mass-loss assumption
of Equation (\ref{equation_dmdt}) the ``super-Eddington wind,''  
referring to Shaviv (2001) \cite{sha01}.  
This interpretation is incorrect.
Shaviv's \cite{sha01} super-Eddington idea is that the photospheric
luminosity could greatly exceed the Eddington luminosity of 
the electron-scattering opacity when the opacity is effectively 
greatly reduced in a porous atmosphere. 
This idea is based on the assumption that all the radiative region of the 
envelope is highly inhomogeneous (porous). 
Such an envelope is very different from the nova envelope  
calculated with the MESA code.  Since the luminosity of the MESA code
is based  on the diffusion approximation and spherically symmetry,  
the diffusive luminosity does not exceed
the Eddington luminosity of the electron-scattering opacity 
at the photosphere as seen in Figure \ref{opac}b. 
Thus, the assumption of Equation (\ref{equation_dmdt}) of the MESA 
code has no theoretical background of the ``super-Eddington wind.''

There are several variations of the method called 
``super-Eddington wind.''
Ma et al. (2013) \cite{ma13} adopted ($L-L_{\rm acc}$) 
instead of $L_{\rm Edd, s}$ in
Equation (\ref{equation_dmdt}), where $L_{\rm acc}=GM \dot M_{\rm acc} /R$.
Wang et al. (2015a) \cite{wan15a} used ($L-L_{\rm acc}^*$), 
where $L_{\rm acc}^*=GM_{\rm WD} {\dot M}_{\rm acc}/R_{\rm WD}$.
Wu et al. (2017)\cite{wu17} replaced $v_{\rm esc}$ 
by $v_{\rm esc}^*\equiv \sqrt{2GM_{\rm WD}/R_{\rm WD}}$,
even though $R$ is nearly two orders of
magnitude larger than $R_{\rm WD}$ when the wind is present.
As the envelope matter already expands to the radius $R$, 
this assumption overestimates the escape velocity and, thus, underestimates 
the mass-loss rate by up to two orders of magnitudes.

These different definitions result in different mass-loss rates, which
directly affect the retention efficiency of mass-accreting WDs.
It has been poorly studied whether these mass-loss assumptions are realistic or not.
An incorrect assumption would lead to incorrect results.  
We encourage authors to publish the temporal change in the mass-loss rate,
surface temperature, and internal structure.
Finally, we note that as the nova calculation is not easy,
computer code must not be used as a black box.

\subsection{Roche-lobe overflow}\label{sec_RL}

Cassisi et al. (1998) \cite{cassi98} and Piersanti et al. \cite{pie99,pie00} claimed that
WDs barely grow to the Chandrasekhar mass limit because
the frictional process should be very effective and most of the envelope
is instantly ejected during nova outbursts.
They did not examine the possible effects of the evolutional
change by the secondary star. 
The companion's gravity tends to suppress mass ejection from the WD. 
The frictional effects inject heat and expand the envelope 
at the companion orbit and reduce the density of the nova envelope thereafter. 
The frictional effect is reduced in turn. 
Especially in a recurrent nova, the envelope is thin because of the
very small envelope mass and the frictional effects can be very weak 
(Kato \& Hachisu \cite{kat91a,kat91b}).
Once the optically thick wind is present, 
the frictional effects are very small (see section 5
in Kato \& Hachisu 1994 \cite{kat94h}).

Wolf et al. (2013) \cite{wol13} calculated nova outbursts with the MESA code 
assuming a Roche-lobe overflow using the OPAL opacity.
They eliminate any mass beyond the Roche lobe radius, i.e.,
the photospheric radius does not exceed the Roche lobe radius.
They also neglected the frictional effects owing to the companion motion 
and gravity mentioned above.
Their HR diagram shows a period of super-Eddington luminosity (upward excursion)
with no particular explanation of the reason.
We suppose two possible reasons: one is a mistake in their plotting program,
the other is a numerical problem such as a failure of energy conservation.
There is an erratum report for this work \cite{wol14}, but
no description has been provided concerning this unknown phenomenon. 

The assumption of mass loss due to the Roche lobe overflow has not yet been 
confirmed in nova outbursts. If it is effective, 
the photospheric radius should remain at the Roche-lobe radius, which 
maintains a high photospheric temperature in close binaries 
until the super-soft X-ray phase begins.
This property is inconsistent with nova observations as mentioned above.
Moreover, nova light curves follow the universal decline law in both 
types of binaries: short (e.g., U Sco) and long (e.g., RS Oph)
orbital periods. 
In other words, the nova light curve is well explained by the optically
thick wind theory, which seems to nominate 
the opacity as the major acceleration source, not frictional effects.

\subsection{Connecting with steady-wind} \label{sec_steadywind}

In the pioneering work of nova outbursts, 
Prialnik (1986) \cite{pri86} calculated one nova outburst cycle
and demonstrated that the ``mass-loss rate'' (Equation (\ref{eq:dmdt}))
varies with the radius in the early phase
but becomes quite constant throughout the envelope
in the later phase.
This implies that the steady-state approximation is good in the later phase.

Using this property, Kato and her collaborators calculated
a series of nova light curves based on the sequence of steady-state solutions
of optically thick winds. The mass-loss rate gradually decreases from the
maximum value (i.e., corresponding to the
optical peak), and the effective temperature gradually rises with time.
This steady-state approach shows a good agreement with the observation 
for a number of novae, as explained in Section \ref{sec_ml}. 
This approach, however, may not be applicable to early phases of the outburst.

Kovetz (1998) \cite{kov98} first presented a calculational method by 
which steady-state optically-thick wind solutions are incorporated
in the stellar evolution code.
The author described the equations and surface boundary conditions in detail
but did not present the interior structure nor the exact connecting 
point between the steady-state winds and interior hydrostatic structure.
In his calculation, the wind mass-loss rate
suddenly increases to the maximum value followed by a gradual decrease.
The theoretical visual magnitude maintains a constant value for some time, 
followed by a gradual decrease.

Idan et al. (2013) \cite{ida13} used a calculation code 
based on Prialnik (1986) 
\cite{pri86} together with the mass-loss algorithm 
proposed by Kovetz (1998) \cite{kov98}.
A sufficiently large number of mass grids ($ > 8000$ at the peak) are adopted.
The authors, however, focused on describing the deep interior structures and
did not show the surface regions including the connecting point.

Kato et al. (2017a) \cite{kat17sha} presented a self-consistent method 
by which they fitted the evolution
calculation with the optically thick wind as a surface boundary condition.
They calculated complete nova outburst cycles on 1.2 and 1.38 $M_\odot$ WDs.
The essential difference to the method in Section \ref{sec_subtract} 
is to set an additional boundary condition of the critical point \cite{kat94h} that 
is a representative boundary condition of steady-steady winds.   
Thus, the mass loss rate is determined as an eigenvalue of 
the boundary-value problem of differential equation.  
This method is complicated, requires many iteration cycles, and
requires careful choice of the fitting point because the wind mass-loss rate
is sensitive to it.

\section{Assumptions}\label{sec_assumption}

\subsection{Initial WD temperature} \label{sec_WDT}

Prialnik \& Kovetz (1995) \cite{pri95} and Yaron et al. (2005) \cite{yar05} 
presented a table of hydrogen shell flashes for three parameters, 
the WD mass, mass accretion rate, and central temperature of the WD.
They called this ``the grid model'' 
in the $(M_{\rm WD}, \dot M_{\rm acc}, T_{\rm c})$ space.
In a long-term evolution of a mass-accreting WD 
toward an SN Ia, these three parameters
are not independent of each other. The WD temperature gradually 
increases/decreases, adjusting to heat balance between the energy loss 
by radiative flux and neutrino loss and the energy gain by accretion. 
A WD increases (decreases) 
in temperature for a higher (lower) mass accretion rate.

Epelstain et al. (2007) \cite{epe07} calculated 1000 successive hydrogen
shell flashes on a 1.0 $M_\odot$ WD with a mass accretion rate of 
$1\times 10^{-11}M_\odot$~yr$^{-1}$ and initial WD temperature of 
$T_{\rm c}=3 \times 10^7$ K at the center.  
After about 220 cycles ($t=6.9 \times 10^8$ yr), $T_{\rm c}$ 
decreased by a factor of 4, the mass-loss period duration
increased by a factor of 4, and the recurrence period increased 
by a factor of 12. If the initial WD is very cold, 
the shell flash would be stronger and a larger part of the accreted matter
would be ejected. As a result, the mass increasing rate of 
the WD would be small.

The authors also showed a 0.6 $M_\odot$ WD
with $\dot M_{\rm acc}=1.0 \times 10^{-9}M_\odot$~yr$^{-1}$. 
An initially hotter WD ($T_{\rm c}=5 \times 10^7$ K) decreases
in temperature, whereas an initially cooler ($5 \times 10^6$ K) WD
increases in temperature.  The WD temperatures finally approach 
a common equilibrium value after 3000 cycles.
Hillman et al. (2016) \cite{hil16} also presented a similar phenomenon
in successive helium shell flashes, in which the central temperature
increased by a factor of 5 through 400 weak helium shell flash cycles
($t\sim 6 \times 10^6$ yr).

In this approach, if we start the calculation from an arbitrarily 
chosen WD temperature,
we must calculate a large number of successive shell flashes
until the WD reaches thermal equilibrium.
During a long-term evolution with accretion, 
the WD temperature may be adjusted
to the thermal equilibrium temperature unless the mass accretion rate changes
more rapidly than the evolution time.
For example, in Hillman et al. (2016), the central temperature 
increased from $1 \times 10^7$ K to $\sim 5.5 \times 10^7$K in a timescale of 
$t\sim 6 \times 10^6$ yr. Nomoto \& Iben (1985) \cite{nom85} showed that the 
central temperature of $1.0~M_\odot$ WD with $\dot M_{\rm acc}=2 
\times 10^{-6}~M_\odot$ yr rises 
from $1 \times 10^7$ K to $> 1 \times 10^8$ K after $2 \times 10^5$ yr. 
In the binary evolution toward an SN Ia explosion like 
in Figure \ref{nomotoD}, the typical mass accretion rate is somewhere
between the above two. Thus,  
the WD temperature may reach equilibrium in a shorter time  
than the evolution timescale to an SN Ia ($\sim 10^7$ yr).

Therefore, an efficient way to save CPU time is to start with 
a WD in thermal equilibrium with a given mass accretion rate.
Townsley \& Bildsten (2004) \cite{tow04} obtained such WD temperatures
in an equilibrium state.
For example, the WD temperature of 1.0 $M_\odot$ is
$T_{\rm c} \sim 4 \times 10^6$ K
for $\dot M_{\rm acc}=10^{-11}~M_\odot$yr$^{-1}$
and $T_{\rm c} \gtrsim 1\times 10^7$ K for $10^{-8}~M_\odot$yr$^{-1}$.
Kato et al. (2017b) \cite{kat17shb} calculated 1500 successive
hydrogen flashes starting from the WD in thermal equilibrium 
and demonstrated that the central WD temperature barely changes 
from the initial value ($\log T_{\rm c}$ (K) = 8.0299) to
the final value (8.0304). The recurrence period soon (after 70 cycles)
approaches the final value after only a small amplitude (8 \%) of variation,
this difference of which is due to the initial envelope structure
that is slightly different from those at later times. 
Wu et al. (2017)\cite{wu17} calculated long term evolution 
of successive He flashes starting from $1.0~M_\odot$ WD 
with the initial WD temperature of $\sim 10^8$ K until it reaches 
$1.378~M_\odot$. Piersanti et al. (2014) \cite{pie14} 
adopted initially cool and hot WD 
models of central temperature $\log T_{\rm c}~({\rm K}) \sim 7.8 - 7.97$. 
These temperatures are out of range of the grid model 
presented by Prialnik \& Kovetz (1995) \cite{pri95} 
and Yaron et al. (2005) \cite{yar05} 
($T_{\rm c}$=1, 3, and 5 $ \times 10^7$ K).
Thus, one should be careful to use their tables for the long-term
evolution of recurrent nova systems.

\subsection{Novae above the stability line: forced novae}
\label{section_stability}

In Section \ref{sec_introduction}, we explain that
hydrogen nuclear burning is stable and no nova outbursts occur in
the region above the stability line 
($\dot M > \dot M_{\rm stable}$ in Figure \ref{nomotoD}).
However, several groups have calculated periodic shell flashes in this region
\cite{pri95,yar05,ida13,hil15}.
Especially, Starrfield et al. (2012) \cite{star12}, 
Idan et al. (2013) \cite{ida13}, 
and Hillman et al. (2015) \cite{hil15}
claimed that there is no steady-state burning,
because they obtained unstable shell flashes in
their calculations.

Hachisu et al. (2016) \cite{hac16forced} clarified the reason 
for this discrepancy. Periodic shell flashes could be
obtained only if accretion is stopped during a flash and restarted
after hydrogen burning ends.  
The trick was in the manipulated on/off switch of accretion. 
If they continued accretion after the first flash,
they would obtain steady-state burning (see Figs. 6 and 7
of Hachisu et al. 2016 \cite{hac16forced}).
To confirm this explanation, Hachisu et al. calculated a test model
with the same parameters as those in Idan et al. (2013) \cite{ida13} 
and reproduced exactly the same periodic shell flashes (see Figure 5
of Hachisu et al. 2016 \cite{hac16forced}).
Thus, the manipulating mass accretion (on/off switch)
was the reason for their periodicity.
Hachisu et al. also demonstrated that the recurrence period (the flashes)
increases (strengthens) if one restarts the accretion at later times.
In this approach, manipulating the mass accretion enables us to design 
shell flash properties. They call such novae ``forced novae.''

The mass increasing rate above the stability line depends on the assumption of 
forced novae. It is possible for forced novae to exist in nature, 
but we have not yet found the observational counterpart.
Instead, we have persistent SSSs above the stability line,
corresponding to steady burning WDs, as in Figure \ref{sss}.
For the optically thick wind region, 
we have the intermittent supersoft X-ray phase 
binary, as explained in Section \ref{sec_introduction}.

Starrfield et al. (2004) \cite{star04} calculated accreting WDs of 
1.25 and $1.35~M_\odot$ with various mass accretion rates and 
concluded that all the hydrogen burning is stable, thus, no shell 
flashes occur at all. This conclusion is clearly incorrect because we have
nova outbursts in the real world, which are thermonuclear runaway events.
Nomoto et al. (2007) \cite{nom07} criticized the paper and clarified 
the reason this incorrect conclusion was reported: because of too small a
grid number, the outermost mass grid was allocated to a mass larger 
than the hydrogen ignition mass, so a shell flash could not occur.
After this criticism, Starrfield et al. (2012) \cite{star12} recalculated the 
nova outbursts with a sufficiently small mass for the outermost grid
and concluded that nuclear burning is unstable
in all the cases (for 0.4--1.35 $M_\odot$ with
$\dot M_{\rm acc}=1.6\times 10^{-11}$ -- $1.6\times 10^{-6}~M_\odot$yr$^{-1}$)
and no steady burning exists. 
This is again inconsistent with the observation of
permanent SSSs, which is considered to be a
``steady burning'' object in Figure \ref{nomotoD}.
After that, the Starrfield group revised their conclusion
based on the calculation with the MESA code (Newsham et al. (2014)
\cite{new14}); they obtained similar results as Figure \ref{nomotoD}, but
the upper region is denoted as ``red giants'' instead of ``wind evolution.''
This was the understanding during the 1980s before the OPAL opacity appeared.
At that time, people thought that the WD becomes a red giant 
(Nomoto diagram 1982) \cite{nom82}.
Currently, the optically thick wind is accelerated by the opacity peak, 
and the ``red giant-like expansion'' is replaced 
by ``optically thick winds.''

\section{On the long-term evolution of a WD and mass retention efficiency}
\label{section_assumption}

There are many issues that may cause different results 
in the long-term evolution of mass-accreting WDs. 
Here, we summarize the main problems in relation to 
the mass retention efficiency. 

\bigskip
\noindent
1. Opacity\\
Opacity has a very important role in mass ejection from mass-accreting WDs. 
If we use OPAL or OP opacities, optically thick winds are always present when 
the condition is satisfied (see Section \ref{sec_opacity}).
Therefore, all the SD and DD scenarios that have been calculated with 
the old opacity (e.g., \cite{whe73,ibe84,ibe96}) should be 
reconstructed by incorporating the new opacity.
Modern SD scenarios based on the OPAL opacity have been proposed
by Hachisu and his collaborators, including optically thick winds
as an elementary process \cite{hkn99,hknu99}.  
The progenitor systems of the SD scenario evolve along with
accretion wind evolution, steady burning, and recurrent nova outburst
phases, the existence of which are supported by observational counterparts
(see Section \ref{sec_nomotodiagram}). 

\bigskip
\noindent
2. The number of mass shell grids\\
In calculating the long-term evolution of a mass-accreting WD, 
it is necessary to adopt a sufficiently large number of mass grids.
Readers should carefully check whether the results might be very different
because of the small number of mass grids. 
(see Section \ref{sec_masszone} and Table \ref{table_mesh}).

\bigskip
\noindent
3. Initial WD temperature\\
The WD temperature is not a free parameter in the long-term evolution, 
because a mass-accreting WD is in thermal equilibrium, 
adjusting with its mass accretion rate. If we adopt a colder WD, 
shell flashes would be stronger and the mass increasing rate 
would be much smaller than the realistic case (see Section \ref{sec_WDT}).

\bigskip
\noindent
4. Mass-loss algorithm \\
One of the most difficult and essential problems in the numerical calculation 
is the mass-loss algorithm.
The most plausible mass-loss mechanism is the optically thick wind
because it is very consistent with nova observations. 
A calculation method combining an evolution code with an
optically thick wind solution has recently been presented 
in two cases of recurrent novae, but it still needs to be improved 
(See Discussion in \cite{kat17sha}).

\bigskip
\noindent
5. Novae above the stability line\\
Shell flashes above the stability line are forced novae. Its mass 
increasing rate depends on the adopted on/off time in the manipulated
mass accretion and is very different from the standard picture 
of ``steady burning'' and ``accretion wind evolution''
(see Section \ref{section_stability}).

\bigskip
\noindent
6. Uncertainty of mass retention efficiency under the stability line\\
Kato et al. (2017b) \cite{kat17shb} first obtained the mass-loss rate
during a full hydrogen flash cycle for a 1.38 $M_\odot$ WD 
with $\dot M_{\rm acc}=1.6 \times 10^{-7}~M_\odot$yr$^{-1}$
($P_{\rm rec}=0.91$ yr), consistent with the observation of 
M31N 2008-12a.  The obtained mass retention efficiency was $\eta =0.4$,
which should be taken as a lower limit because they neglected rotation
effects as well as other effects.
For He shell flashes, no time-dependent calculation has succeeded in
incorporating optically thick winds. 
Thus, the mass increasing rates of WDs remain to be improved  
below the stability line. 
A better approach, so far, may be those of the steady-state 
sequence calculated by Kato and her collaborators. 

\bigskip
\noindent
7. Unsolved problems\\ 
The numerical calculation listed in Table \ref{table_mesh} are all 
one-dimensional calculations.  They do not include the effects of rotation
or dredge-up of WD material due to shear mixing at the boundary layer. 
Yoon et al. (2004) \cite{yoo04} showed that rotation generally makes 
flashes milder because of the weaker gravity as well as the contamination 
of C/O by rotational mixing at the base of the He layer. Kato et al. (2017b) 
\cite{kat17shb} discussed that it is possible for mass increasing 
rate to increase but did not provide a quantitative estimate.
Thus, the real mass retention efficiency remains to be improved. 

\bigskip
\acknowledgments
 This research has been supported in part by the Grants-in-Aid for
 Scientific Research (15K05026, 16K05289)
 of the Japan Society for the Promotion of Science.

\bigskip
\bigskip
\noindent{\bf DISCUSSION}

\bigskip
\noindent{\bf J. Isern}: What is the role of rotation, if any, 
in the frequency and intensity of the H and He flashes?

\bigskip
\noindent
{\bf Kato}: The effects of rotation have not been fully studied yet. 
The 1D calculation of He flashes by Yoon et al. (2004) shows that 
the recurrence period (and thus intensity of the flash) 
does not change, but the He mass retention efficiency increases.  

\bigskip
\noindent
{\bf G. Shaviv}: (comments at the end of the session) 
You criticized the small mass-grid number in the numerical calculations. 
I agree with that, but, in addition to the total number of mass grids, 
you should also allocate enough grids where the gradient of physical
values is high.

\bigskip
\noindent
{\bf Kato}: I added the warning by Nariai et al. (1980) \cite{nar80}
in Section \ref{sec_masszone}. In any case, it is clear that 
only 5 mass grids (after 40 cycles) in the He-rich layer 
in Figure \ref{QXm135}b are not enough.  

\end{document}